\documentclass[prb,preprint,draft,amsmath,showpacs]{revtex4}

\usepackage{graphicx}
\begin{document}

\DeclareGraphicsExtensions{.eps, .jpg}

%%%%%%%%%%%%%%%%%%%%%%%%%%%%%%%%%%%%%%%%%%

\bibliographystyle{prsty}
\input epsf

\title {Infrared signature of the charge-density-wave gap in $ZrTe_3$}

\author {A. Perucchi$^{1}$, L. Degiorgi$^{1}$, and H. Berger$^{2}$}
\affiliation{$^{1}$Laboratorium f\"ur Festk\"orperphysik, ETH Z\"urich,
CH-8093 Z\"urich, Switzerland}\
\affiliation{$^{2}$Department of Physics, EPF Lausanne,
CH-1015 Lausanne, Switzerland}\

\date{\today}

\begin{abstract}
 The chain-like $ZrTe_3$ compound undergoes a charge-density-wave (CDW) transition at $T_{CDW}=63$ $K$, most strongly affecting the conductivity perpendicular to the chains. We measure the temperature ($T$) dependence of the optical reflectivity from the far infrared up to the ultraviolet with polarized light. The CDW gap $\Delta(T)$ along the direction perpendicular to the chains is compatible for $T<T_{CDW}$ with the behavior of an order parameter within the mean-field Bardeen-Cooper-Schrieffer (BCS) theory. $\Delta(T)$ also persists well above $T_{CDW}$, which emphasizes the role played by fluctuation effects.
\end{abstract}
\maketitle

%\section{Introduction}
The physical properties of low-dimensional layered systems, like the transition metal trichalcogenides ($MX_3$, $M=Nb, Ta$ and $X=Se, S$) conductors, received considerable attention even before the discovery of high-$T_c$ superconductivity in the layered copper oxides \cite{rouxel}.  Most recently, the semi-metallic $ZrTe_3$ also acquired renewed interest because of its peculiar transport properties. Its crystal structure (the so-called  B variant) consists of infinite rods formed by stacking $ZrTe_3$ prisms along the chain b-direction. There are two identical chains connected by inversion symmetry in the monoclinic unit cell \cite{felser}, so that in $ZrTe_3$ the two neighboring chains, arranged in sheets parallel to the transverse a-direction, are alternate chains pair. In view of this crystal structure, one might assume that, like in a typical $d$ band metal as $NbSe_3$, the electrical transport properties are associated with the regularly spaced metal atoms along the trigonal-prism chain. However, $ZrTe_3$ undergoes a charge-density-wave (CDW) transition at $T_{CDW}=63$ $K$, which, quite astonishingly, most strongly affects its conductivity components perpendicular to the conducting chains \cite{takahashi}. The electrical resistivity ($\rho(T)$) of $ZrTe_3$ is anisotropic ($\rho_b\sim\rho_a\sim\rho_c/10$) and points to a two-dimensional (2D) behavior. $ZrTe_3$ remains metallic below the bump anomaly in $\rho(T)$ along the directions perpendicular to the chain axis, while along the b-axis \cite{takahashi} it only presents an anomaly at $\sim 55$ $K$ in the first derivative of $\rho_b$. Furthermore, $ZrTe_3$ displays filamentary superconductivity below 2 $K$ (Ref. \onlinecite{takahashi}).

In the CDW state, the formation of electron-hole pairs with wave vector $q_{CDW}$, connecting (nesting) one large portion of the Fermi surface (FS) with another, leads to the opening of a gap. The formation of a FS energy gap is a fundamental quantum phenomenon in solids, because thereby the system of interacting electrons can stabilize a broken symmetry ground state. The precise measurement of the (CDW) gap permits furthermore a meaningful comparison to microscopic theories \cite{grunerbook}.
The residual metallicity below $T_{CDW}$ is due to the deviation from perfect nesting, which leaves pockets of itinerant carriers \cite{comment}.  $ZrTe_3$, like other spin and charge density wave systems (as $Cr$, $NbSe_3$ or $CeTe_3$, Ref. \onlinecite{grunerbook}), is then of interest for studying the impact of the CDW collective state on the metallic properties.

We provide here the first comprehensive study of the anisotropic optical response of $ZrTe_3$ over a broad spectral range and as a function of temperature ($T$). Previous optical spectroscopy measurements \cite{khumalo, heer, bayliss} were performed at 300 $K$ or over an insufficient energy range, to establish the relevant energy scales associated with the collective CDW state. Our optical conductivity along the direction perpendicular to the chains shows a redistribution of spectral weight from low to high frequencies with decreasing $T$. We determine the bulk CDW gap \cite{comment}, which turns out to greatly follow the BCS behavior of an order parameter for $T<T_{CDW}$. The gap feature already develops at $T$ above $T_{CDW}$. This indicates that the precursor effects of the CDW phase transition play a relevant role. Such fluctuation effects strongly influence the normal state properties. The excitation spectrum of $ZrTe_3$ is quite reminiscent of that of complex materials exhibiting pseudo-gaps in some directions, as in certain phases of cuprates \cite{reviewopticshtc}.

$ZrTe_3$ single crystals were prepared by chemical vapor transport. An almost stoichiometric mixture of powdered $Zr$ and $Te$, having a slight excess of $Te$, was enclosed in an evacuated and sealed quartz ampoule along with iodine as the transport agent. 
Charge and growth-zone temperatures were 800 $^{\circ}C$ and 750 $^{\circ}C$, respectively. 
The crystals  have flat large surfaces ($\sim$ 2 $mm$ x 5 $mm$), and are elongated along the chain b-axis. We measure the $T$ dependence of the optical reflectivity $R(\omega)$ from the far infrared up to the ultraviolet (i.e., between 30 and $10^5$ $cm^{-1}$, Refs. \onlinecite{wooten} and \onlinecite{dressel}). Light is polarized along the chain (E$\parallel$b) and along the direction perpendicular to the chain (E$\perp$b), in order to assess the anisotropic electrodynamic response. Through Kramers-Kronig transformation we extract the real part $\sigma_1(\omega)$ of the optical conductivity, representing the complete absorption spectrum. To this end, we employ standard high-frequency extrapolations $R(\omega)=\omega^{-s}$ (with $2<s<4$) in order to extend the data set above $10^5$ $cm^{-1}$ and into the electronic continuum. $R(\omega)$ is extrapolated towards zero frequency using the Hagen-Rubens (HR) law $R(\omega)=1-2\sqrt{(\omega/\sigma_{dc})}$ from data points in the 30-70 $cm^{-1}$ range. The $T$ dependence of $\sigma_1(\omega)$ is not affected by the details of this low-frequency extrapolation \cite{commentHR}.

Our optical $R(\omega)$ spectra (Fig. 1) display an overall metallic behavior at any temperatures, and an anisotropic behavior between the two polarization directions is observed up to $\sim 40000$ $cm^{-1}$. We note, however, that the anisotropy is much less pronounced with respect to the transition metal trichalcogenides  $NbSe_3$ (Ref. \onlinecite{perucchi04_Nb}) and $TaSe_3$ (Ref. \onlinecite{perucchi04_Ta}). This is due to the increased role played  by the $p$ bands in $ZrTe_3$, if compared to the selenide systems, where the metallic ($d$) channel along the b-axis dominates. Our $R(\omega$) measurement shows the presence of two strong ``dips'' at $\sim 2000$ and 4000 $cm^{-1}$ which are not observed in previous optical measurements \cite{khumalo, heer, bayliss}. The reason for these differences is not known at present. The rise of $R(\omega)$ below 2000 $cm^{-1}$ defines the $R(\omega)$ plasma edge, and $R(\omega)$ merges to total reflection for $\omega\rightarrow0$. A blow-up of the spectra in the far infrared spectral range for the more interesting E$\perp$b polarization is shown in the inset of Fig. 1. 

%<<<<<<<<<<<<<<<<<<<<<<<< FIGURE>>>>>>>>>>>>>>>>>>>>>>>>>
\begin{figure}[h]
   \begin{center}
   %\resizebox{9.0 cm}{!}{\includegraphics{Figura1.epsf}}
    \leavevmode
    \epsfxsize=7cm \epsfbox {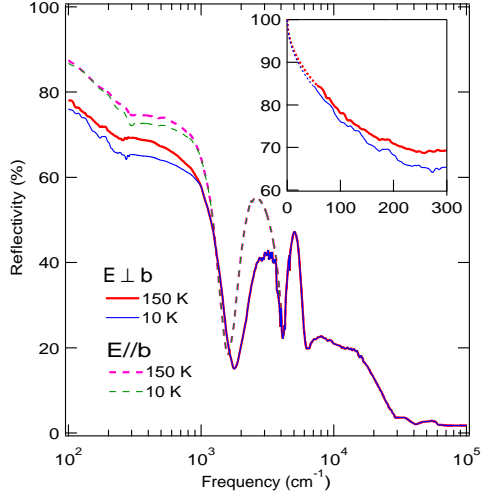}
     \caption{Reflectivity spectra $R(\omega)$ of $ZrTe_3$ for both polarization directions (E$\parallel$b and E$\perp$b) from the far infrared up to the ultraviolet. Inset: $R(\omega)$ of $ZrTe_3$ in the far infrared range at 150 and 10 $K$ for the E$\perp$b polarization. The dotted line is the low frequency HR extrapolation for $\omega<30$ $cm^{-1}$.}
\label{fig1}
\end{center}
\end{figure}
%<<<<<<<<<<<<<<<<<<<<<<<< figure >>>>>>>>>>>>>>>>>>>>>>>>>

The effective metallic component is not a simple Drude. This is better seen in $\sigma_1(\omega)$ (Fig. 2a): besides the finite optical conductivity for $\omega\rightarrow0$, characteristic for the metallic behavior, a broad maximum at about 400 $cm^{-1}$ overlaps the low frequency metallic component. The feature at 400 $cm^{-1}$ is strongly asymmetric and presents a large tail towards high frequencies. 
Several absorptions (inset Fig. 2a) are recognized above $10^4$ $cm^{-1}$, which are ascribed to $p-d$ electronic interband transitions \cite{bayliss}. The absorptions at 400, 3230 and 4840 $cm^{-1}$ (inset Fig. 2a) bear some similarities with band features recently observed by ARPES experiments \cite{grioni}. The feature at 400 $cm^{-1}$ may be compatible with an electron pocket at the $B$-point of the Brillouin zone (BZ), associated with a saddle point characterized by a large density of states. The excitation at 3230 $cm^{-1}$ may be related to a strong maximum, due to a big hole pocket, at the $\Gamma$-point of BZ.  At the $B$-point, the ARPES spectra also suggest the presence of a minimum in the electronic band structure, which could lead to the electronic optical transition at 4840 $cm^{-1}$ (Ref. \onlinecite{grioni}). 

%\section{Experiment and results}

By decreasing $T$ down to 10 $K$, we clearly observed a depletion of  $R(\omega$) below $\sim 1000$  $cm^{-1}$ (Fig. 1). Such a depletion is more pronounced in the case of the E$\perp$b polarization. Furthermore, an enhancement of tiny infrared features (particularly at 180 and 270 $cm^{-1}$, inset Fig. 1 and Fig. 2a) is observed with decreasing $T$. These features might be due to the presence of so-called {\it phase phonons} \cite{degiorgi91,perucchi04_Nb,perucchi04_Ta}, arising from the coupling of the CDW condensate with the lattice degrees of freedom \cite{rice76}.

%<<<<<<<<<<<<<<<<<<<<<<<< FIGURE>>>>>>>>>>>>>>>>>>>>>>>>>
\begin{figure}[h]
   \begin{center}
   %\resizebox{9.0 cm}{!}{\includegraphics{Figura2.epsf}}
    \leavevmode
    \epsfxsize=8cm \epsfbox {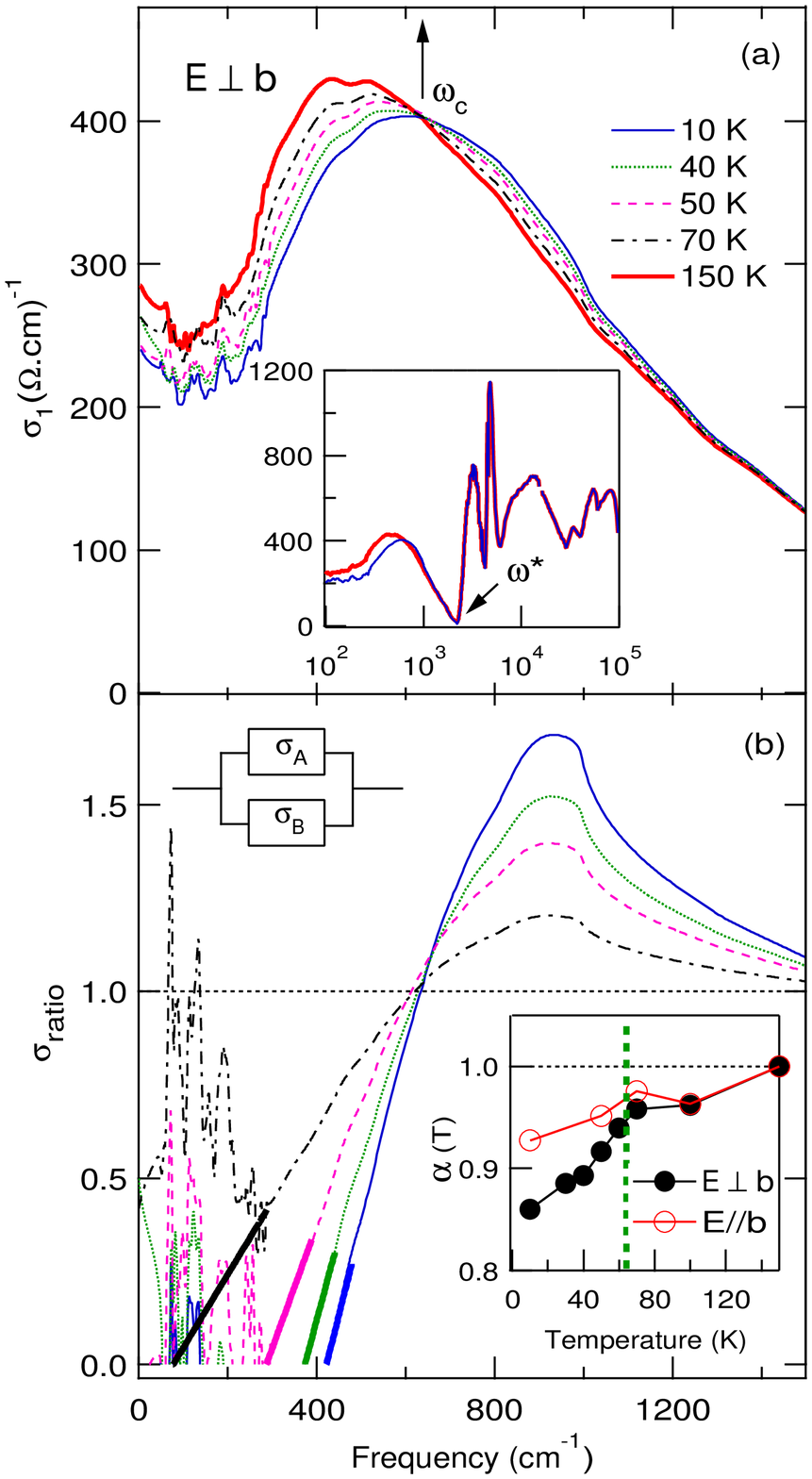}
     \caption{(a) Real part $\sigma_1(\omega)$ of the optical conductivity in the infrared range at selected $T$, for light polarization E$\perp$b. The upwards arrow indicates $\omega_c$, the frequency at which the $\sigma_1(\omega)$ curves cross. Inset: $\sigma_1(\omega)$ at 10 and 150 $K$ (E$\perp$b) from $10^2$ up to $10^5$ $cm^{-1}$. $\omega^*$ (downwards arrow) is the cut-off frequency for the integration in eq. (\ref{firstmoment}). (b) $\sigma_{ratio}$ (see text) at selected $T$ for E$\perp$b. The bold segments indicate the linear extrapolation of $\sigma_{ratio}$ towards the frequency axis at $2\Delta$ so that $\sigma_{ratio}(2\Delta)=0$, defining the CDW gap. Inset: $\alpha(T)=S\omega(T)/S\omega(150$ $K)$, obtained from eq. (1) for both polarization directions\cite{comment2}. The bold dashed line marks $T_{CDW}=63$ $K$ from the resistivity measurement  \cite{takahashi}.}.
\label{fig2}
\end{center}
\end{figure}
%<<<<<<<<<<<<<<<<<<<<<<<< figure >>>>>>>>>>>>>>>>>>>>>>>>>

We now focus our attention on the low frequency range for E$\perp$b and first define $\omega_c\sim$ 630 $cm^{-1}$ as the frequency at which all $\sigma_1(\omega)$ curves cross (Fig. 2a). We then consider the partial sum \cite{dressel}
\begin{equation}
Sw(T)=\int_0^{\omega_c}\sigma_1(\omega,T)d\omega.
\end{equation}
Below $\omega_c$, there is a depletion of $S\omega$ with decreasing $T$, which is transfered at higher energies \cite{comment3}. The redistribution of $S\omega$ is a typical signature for the opening of a gap \cite{comment}, as a consequence of the CDW phase transition. Since $\int \sigma_1(\omega)d\omega\sim \frac{n}{m}$ (where $n$ is the charge carriers density, and $m$ is their effective mass), $\alpha(T)=S\omega(T)/S\omega(150$ $K)$ estimates the portion of FS which remains ungapped \cite{comment2}. The inset of Fig. 2b shows a continuous decrease of $\alpha(T)$, which becomes more pronounced below 70 $K$, i.e. in the proximity of $T_{CDW}$. From our data, we can establish that 86\% of the FS present at 150 $K$ still survives at 10 $K$ for E$\perp$b. This result is in fair agreement with the rough estimation of the FS gapping from the anomaly at $T_{CDW}$ in $\rho(T)$, as performed by Ong and Monceau for the transport properties of $NbSe_3$ (Ref. \onlinecite{ong77}). The same analysis in terms of $S\omega(T)$,  applied to the data for E$\parallel$b polarization, leads to qualitatively similar results, even though the $T$ dependence of $\alpha(T)$ is weaker (inset Fig. 2b). The cut-off frequency in eq. (1) was chosen to be $\omega_c\sim 570$ $cm^{-1}$ for E$\parallel$b. A ``kink'' in $\alpha(T)$ is also present at $\sim 63$ $K$, and the FS portion remaining ungapped at 10 $K$ is larger than 90\%. This confirms that  the CDW phase transition along the chain direction is much less affecting the FS of $ZrTe_3$. It is worth reminding that also in the structurally closer $TaSe_3$ compound a large FS pseudo-gapping is observed along the direction perpendicular to the chains \cite{perucchi04_Ta}.

For the E$\perp$b polarization, we assume that $\sigma_1(\omega)$ can be described by two parallel conducting channels ($\sigma_1=\sigma_{1A}+\sigma_{1B}$, Fig. 2b).  We further postulate that these two channels have the same frequency dependence in the normal state (i.e., $\sigma_{1A}^N(\omega)=\beta\cdot\sigma_{1B}^N(\omega)$, $\beta$ being a constant), but one of them ($\sigma_{1A}$) is affected by the CDW phase transition whereas the other one ($\sigma_{1B}$) remains completely ungapped at low $T$. Therefore, since ($\sigma_{1B}^{CDW}=\sigma_{1B}^N$), we can write
\begin{equation}
\frac{\sigma_1^{CDW}}{\sigma_1^N}=\frac{\sigma_{1A}^{CDW}+\sigma_{1B}^N}{\sigma_{1A}^{N}+\sigma_{1B}^N},
\end{equation}
which corresponds to the well known Mattis-Bardeen representation \cite{tinkham} of the absorption spectrum in the CDW state. It can be shown that our original assumption also leads to $\sigma_{1B}^N/\sigma_1^N=1/(1+\beta)=\alpha$. We can then define 
\begin{equation}
\sigma_{ratio}(\omega)=\frac{\sigma_{1A}^{CDW}}{\sigma_{1A}^N}=\left(\frac{\sigma_1^{CDW}}{\sigma_1^N}-\alpha\right)\frac{1}{1-\alpha}.
\end{equation}

$\sigma_{ratio}(\omega)$ is illustrated in Fig. 2b for $\alpha=0.86$, as estimation of the ungapped portion of the FS at 10 $K$ (inset Fig. 2b). This procedure permits to clearly discriminate between the effect due to the CDW transition on channel A and the  ``normal'' (i.e., B-channel) background. The findings in Fig. 2b bear a striking similarity with the theoretical predictions for a CDW system within the Lee-Rice-Anderson approach \cite{lee74}. Indeed, the peak in $\sigma_{ratio}$ at 900-1000 $cm^{-1}$ may be ascribed to the CDW single-particle excitation. The onset of the gap-absorption gets steeper with decreasing $T$. Furthermore, we claim that the formation of a peak in $\sigma_{ratio}$ for the CDW gap feature is the consequence of the so-called case I coherence factors \cite{dressel}, as postulated by the BCS theory \cite{tinkham}.   

We extract the $T$ dependence of the CDW gap\cite{comment} by linearly extrapolating (bold segments in Fig. 2b) the $\sigma_{ratio}$ curves to intersect the frequency axis at $\omega=2\Delta$, where $\sigma_{ratio}(2\Delta)=0$. As quite typical for CDW systems \cite{grunerbook}, $2\Delta(T)/k_BT_{CDW}\simeq 9$ is a few times greater than the BCS expected value of 3.52 (Ref. \onlinecite{tinkham}). This would imply a mean field $T_{CDW}^{MF}$ of about 160 $K$. $\Delta(T)/\Delta(10K)$ is plotted in Fig. 3.
An alternative and totally independent procedure to extract the $T$ dependence of the CDW single particle excitation is to quantify the frequency shift of the gap absorption. To this end, we consider the first moment of the contribution due to the gap, defined by the expression \cite{nucara03}
\begin{equation}\label{firstmoment}
<\omega>=\frac{\int_0^{\omega^*}\sigma_1(\omega)\cdot \omega \,  d \omega}{\int_0^{\omega^*}\sigma_1(\omega) \,  d \omega},
\end{equation}
with $\omega^*\sim 2230$ $cm^{-1}$. $\omega^*$ (inset Fig. 2a) was chosen here in order to account for the spectral range within which the $T$ dependence of $\sigma_1(\omega)$ is fully developed. We then calculate the difference $\delta<\omega>(T)=<\omega>(T)-<\omega>(150$ $K)$ (Ref. \onlinecite{comment2}), and plot the ratio $\delta<\omega>(T)/\delta<\omega>(10$ $K)$ in Fig. 3. The fact that the $T$ dependence of $\delta<\omega>$ coincides fairly well with that of $\Delta$, directly extracted from $\sigma_{ratio}$, demonstrates the validity of the overall analysis. In passing, we note that even the determination of the shift in the resonance frequency of the peak at about 400 $cm^{-1}$ in $\sigma_1(\omega)$ (Fig. 2a), extracted within the Lorentz-Drude\cite{wooten,dressel} fit of $\sigma_1(\omega)$, roughly agrees with the behavior found for $\delta<\omega>$ and $\Delta$.

%<<<<<<<<<<<<<<<<<<<<<<<< FIGURE>>>>>>>>>>>>>>>>>>>>>>>>>
\begin{figure}[h]
   \begin{center}
   %\resizebox{9.0 cm}{!}{\includegraphics{Figura3.epsf}}
    \leavevmode
    \epsfxsize=8cm \epsfbox {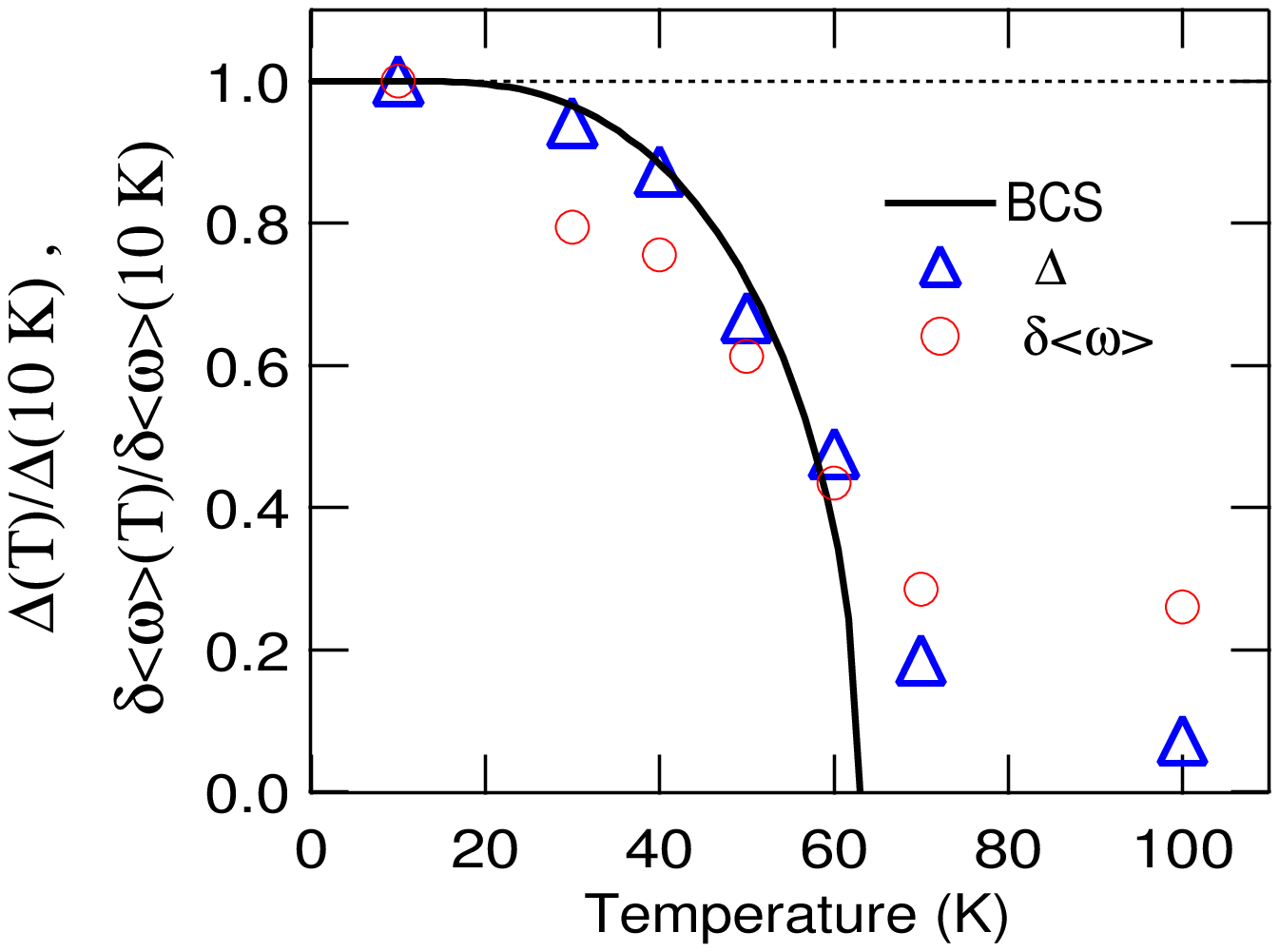}
     \caption{Temperature dependence of the gap values ($\Delta$) as extracted from $\sigma_{ratio}$, and the first moment difference $\delta<\omega>$, normalized by their values at 10 $K$ (see text). The BCS temperature dependence of the order parameter for $T_{CDW}=63$ $K$ (Ref. \onlinecite{takahashi}) is also shown for comparison. }
\label{fig3}
\end{center}
\end{figure}
%<<<<<<<<<<<<<<<<<<<<<<<< figure >>>>>>>>>>>>>>>>>>>>>>>>>
In Fig. 3, we also show  the $T$ dependence of the BCS gap, which mimics the order parameter for the CDW phase transition occurring at the critical temperature $T_{CDW}$. Our data are in fair agreement with the BCS behavior up to $T_{CDW}$. Above $T_{CDW}$, the data deviate from the BCS prediction, indicating that the gap, although strongly reduced, remains open. We claim that CDW fluctuations may explain the behavior of $\Delta(T)$ in the 2D $ZrTe_3$. Precursor effects to the CDW phase transition were already anticipated above by the gradual decrease of $S\omega(T)$ (eq. (1)) at $\omega<\omega_c$ for $T<150 K\sim T_{CDW}^{MF}$ (inset Fig. 2b) and might be compatible with electron microscopy data \cite{eaglesham84}. This suggests that the fluctuation regime extends to $T$ at least twice as large as $T_{CDW}$, which is quite common in CDW systems \cite{schwarz,perucchi04_Nb,johnston84}. The gapping of FS and the corresponding depletion of $S\omega$ at $T_{CDW}$ is then a signature for the crossover into a highly correlated three dimensional state with long range correlation lengths \cite{grunerbook}.

In conclusion, we have presented a detailed optical study on the FS gapping in the anisotropic semi-metallic $ZrTe_3$.  We have proposed a simple procedure for disentangling the role of the gapped and ungapped parts of the FS ($hot$ and $cold$ spots), which may be generalized to other pseudo-gap systems. While behaving as a BCS order parameter, the peculiarity of the CDW single particle gap resides in its development  primarily along the direction perpendicular to the one-dimensional chain. The modulation of the crystal structure along the perpendicular direction is therefore the main ingredient, driving $ZrTe_3$ towards the CDW instability. $ZrTe_3$ provides an ideal ground to test the impact of the CDW phase transition and the FS nesting on the electronic structure. Moreover, our results vividly illustrates the importance of fluctuation effects on the physical properties of low-dimensional conductors. 

%\acknowledgments
The authors wish to thank J. M\"uller for technical help, and M. Grioni, D. Basov, C. S\o ndergaard and G. Caimi for fruitful discussions. This work has been supported by the Swiss National Foundation for the Scientific Research, within the NCCR research pool MaNEP.

\newpage


\begin{thebibliography}{10}
\bibitem{comment} Because of the remaining FS, the CDW single particle excitation should be correctly named pseudo-gap. For simplicity we call it CDW gap.
\bibitem{commentHR}
The HR extrapolation yields $\sigma_{dc}$ values in rough agreement with dc transport data \cite{takahashi}. Nevertheless, a crossing  between $R(\omega)$ at high and low $T$ would be necessary in order to recover the exact $T$ dependence of the metallic dc resistivity. Such a $R(\omega)$ crossing does not yet occur down to 30 $cm^{-1}$.
\bibitem{comment2} We consider the data at 150 $K$ at high enough $T$ with respect to $T_{CDW}$ to be representative for the normal state optical properties.
\end{thebibliography}
\end{document}